\documentclass[11pt]{article}
\usepackage[final]{graphicx}
\usepackage[parfill]{parskip}    
\usepackage{amssymb,amsmath,amsthm}
\usepackage{setspace}
\begin{document}
\title{Emergence of Price Divergence in a Model Short-Term Electric Power Market\footnote{This manuscript has been authored by Sandia Corporation under Contract No. DE-AC04-94AL85000 with the U.S. Department of Energy.  The United States Government retains and the publisher, by accepting the article for publication, acknowledges that the United States Government retains a non-exclusive, paid-up, irrevocable, world-wide license to publish or reproduce the published form of this manuscript, or allow others to do so, for United States Government purposes.}}
\author{Randall A. LaViolette\thanks{ralavio@sandia.gov} \and Lory A. Ellebracht\thanks{Present address: Hallmark Cards Inc., 2500  McGee St., Kansas City, MO, 64108 USA} \and Kevin L. Stamber \and Charles J. Gieseler\thanks{Present address:  RESPEC, 5971 Jefferson NE STE 101, Albuquerque, NM 87109 USA} \and Benjamin K. Cook\\ \\Sandia National Laboratories,\\ P.O. Box 5800, Albuquerque, NM, 87185 USA}
\maketitle

\begin{abstract}
A minimal model of a market of myopic non-cooperative agents who trade bilaterally with random bids reproduces qualitative features of short-term electric power markets, such as those in California and New England.  Each agent knows its own budget and preferences but not those of any other agent. The near-equilibrium price established mid-way through the trading session diverges to both much higher and much lower prices towards the end of the trading session.  This price divergence emerges in the model without any possibility that the agents could have conspired to ``game'' the market. The results were weakly sensitive to the endowments but strongly sensitive to the nature of the agent's preferences and budget constraints.
\end{abstract}
\section{Introduction}
\label{sec:intro}
\begin{figure}[p]
\begin{center}
\includegraphics{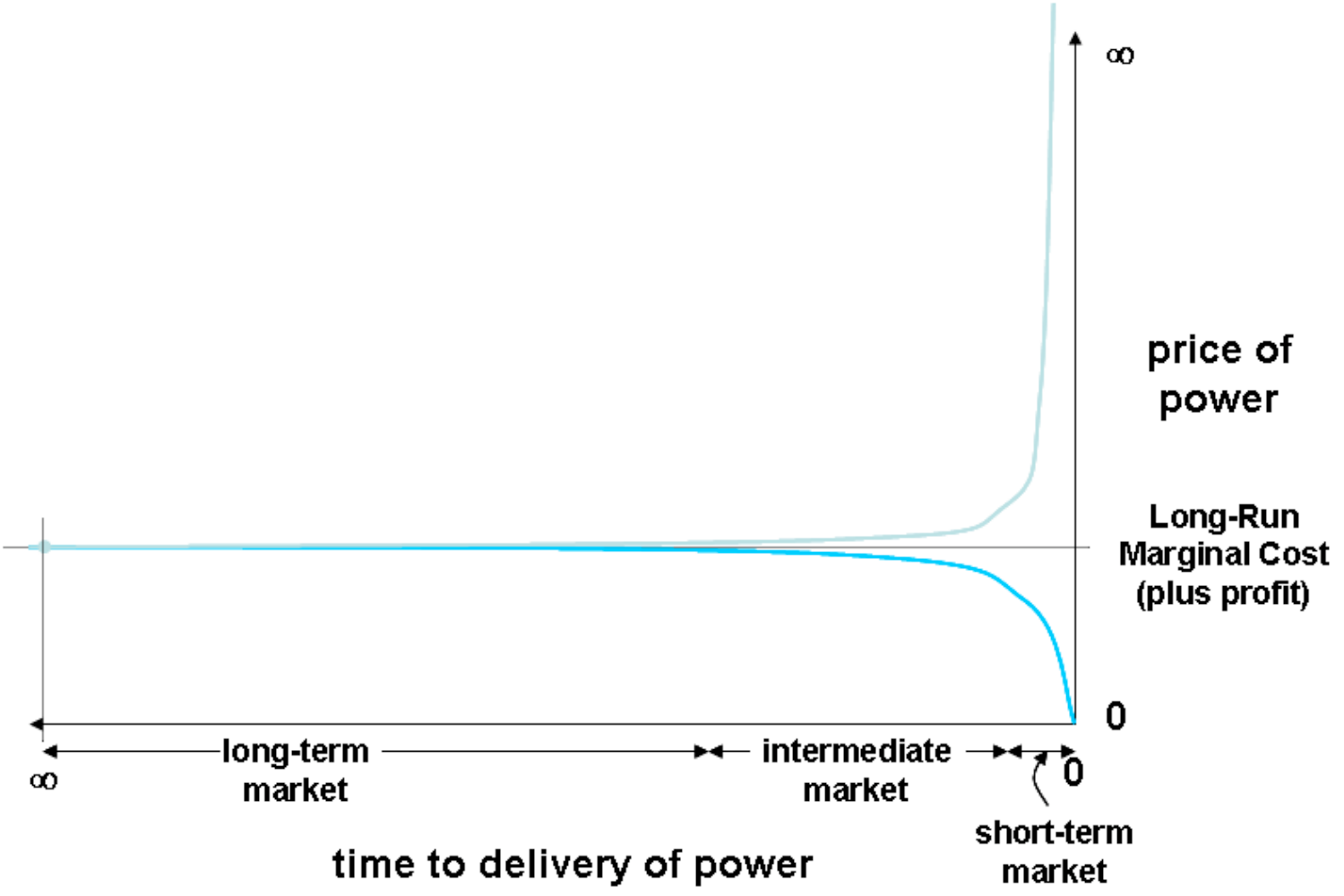}
\caption{Schematic of contracted price as a function of time to deliver power, from \cite{Stamber:2001}.}
\label{fig:Stamber}
\end{center}
\end{figure}

The price evolution in markets for short-term electric power (Figure \ref{fig:Stamber}) contrasts with the textbook examples of two-good market, wherein price fluctuations rapidly diminish as they settle down to an equilibrium price\cite{MICRO:2001a,GodeSunder:1993}.  In the short-term power market the price fluctuations may increase dramatically towards the end of the trading session, resulting in both high and low prices that can be many standard deviations from the mean price set earlier in the session.  This phenomenon could be plausibly attributed to a variety of factors, e.g., the inelasticity of supply and demand or the structure of the market, that impute to the agents extensive market knowledge and strategic skills.  Such assumptions would be consistent with both the recent experience of short-term power markets and with traditional economic 
theory\cite{MICRO:2001a}.

Therefore it may seem surprising that we consider instead the hypothesis that the divergence phenomenon might be attributable merely to random interactions between myopic agents (i.e., agents who know only about themselves) with simple preferences expressed as budget constraints and target quantities of demand or supply.  This hypothesis was inspired by and parallels the work of Gode and Sunder\cite{GodeSunder:1993}, who provided a minimal model of a double-auction market with what they termed ``zero-intelligence'' agents (who bid and ask random prices without regard to the state of the market or other agents) that produced a price evolution that rapidly converged to the same equilibrium price as that predicted by the traditional theory with its assumption of intelligent agents.  Certainly their model did include the one agent endowed with knowledge of the market, i.e., the auctioneer, who would steer the randomly fluctuating prices towards the equilibrium by progressively narrowing the range in which the bids and asks would be accepted.  In the markets of interest here, we consider only bilateral trades without the benefit of a market maker.  
We discuss below a simple modification of the original zero-intelligence assumptions that produce the divergence phenomenon in a bilateral market of myopic agents.

\section{Description of the model}
\label{sec:model}
We assumed that the market consisted of non-cooperative agents who traded bilaterally without the supervision of or input from a market maker.  In the following paragraphs we describe the agents, the initial conditions, and the transaction rules of this model market.  All of the quantities, budgets, or prices were taken to be unitless.  The differences between this model and the classical Edgeworth-box trading paradigm are discussed in the conclusion of this section.  The principal differences between our myopic agent and the classical zero-intelligence agent\cite{GodeSunder:1993} is that our agent possesses multiple units of quantity and that its preferences are expressed through budget constraints rather than with fixed costs or values.  We emphasize that this minimal model does not provide, e.g., a real-time simulation of the transaction dynamics of an actual short-term power market (for which we lack the data in any case); instead, we employed it to test hypotheses about the information agents require in order to obtain the important features of such a market.  

For each trading session we divided the agents into $N_B$ buyers and $N_S$ sellers.  No agent could change roles.  Each agent was endowed with multiple units of integer quantity (supply or demand, see Appendix \ref{appendix:initialization}) chosen randomly from a discrete distribution; the lower bound on the distribution of the quantity was much larger than unity.  We also endowed each agent with an initial budget $B_0$ chosen randomly from a continuous distribution.  The buyer's budget was the amount she could spend to acquire her initial demand $D_0$.  The seller's budget was the amount that he was required to recover from the sale of his initial supply $S_0$.  In contrast to the buyer, his budget was allowed to become negative; in that regime, all sales would contribute to profit beyond his revenue target.  Each agent always knew its own quantity and budget but knew nothing about the budget or quantity of the other agents.  Buyers left the market if their demand was met or if their budget was spent (i.e., they were not allowed to accumulate debt).  Sellers left the market only if their supply sold out.  

\begin{figure}[p]
\begin{center}
\includegraphics[width=16cm]{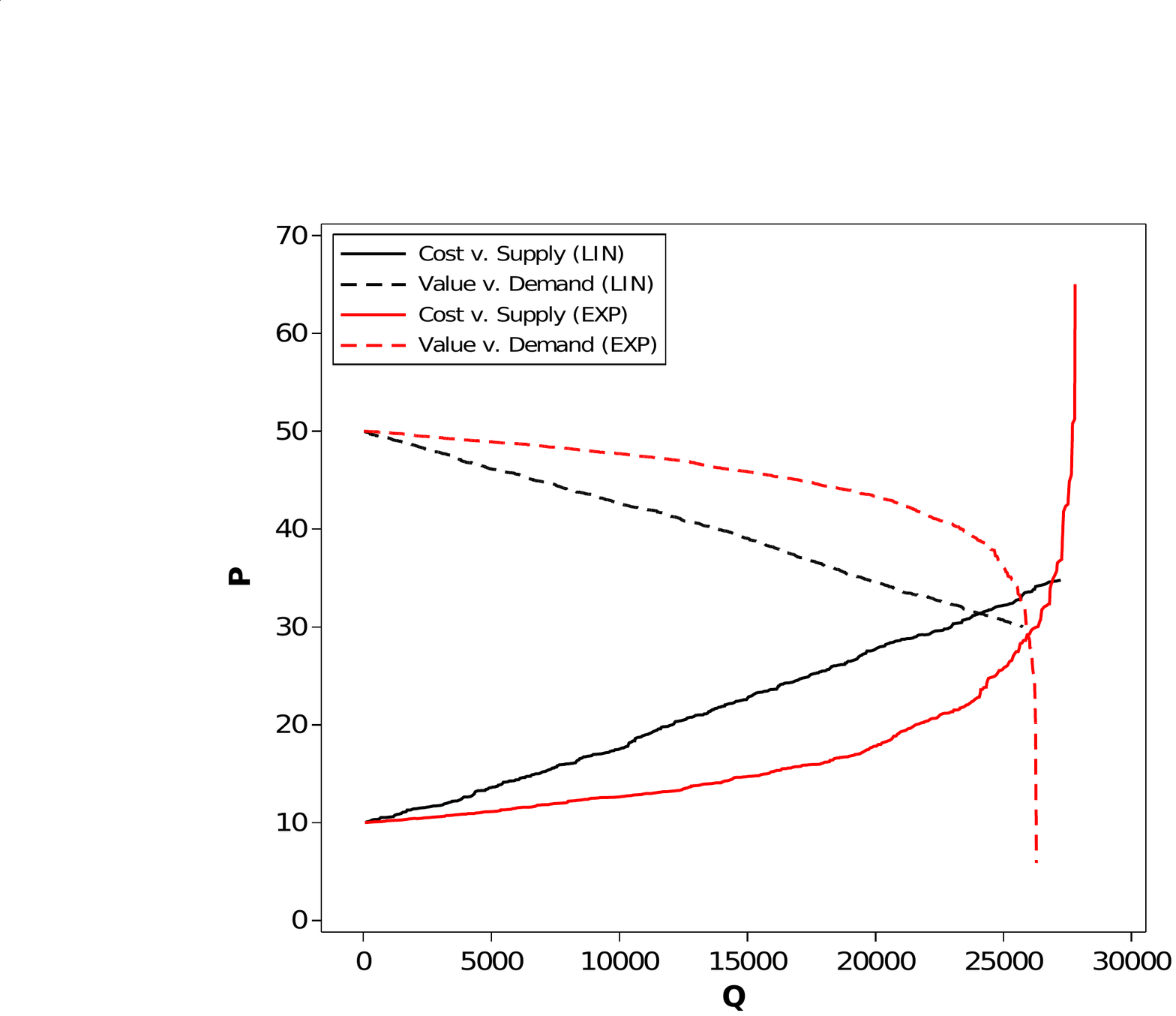}
\caption{Supply-demand curves of the agent's initial conditions for a typical trading session.  See Appendix B for the parameters.  The linear curves (case LIN) are the most elastic where the two curves intersect; the exponential curves (case EXP) are also the least elastic at the intersection.}
\label{fig:supply-demand}
\end{center}
\end{figure}

The initial budget distribution was characterized by an initial aggregate price elasticity parameter $c$ (see  Appendix \ref{appendix:initialization}), which was selected so that the resulting initial budget-quantity curves could appear to be strongly exponential and inelastic (case EXP) or nearly linear and more elastic (case LIN) where they intersected.  We considered these two extreme cases, even though case EXP seems to us to be the more realistic, because we were interested in the sensitivity of the results to different initial conditions.  Figure\ref{fig:supply-demand} shows typical supply-demand curves for the initial conditions, where for each agent its initial value or cost would be interpreted as its $\frac{B_0}{D_0}$ or its $\frac{B_0}{S_0}$, respectively.  We stress that the curves in Figure \ref{fig:supply-demand} applied only to the initial conditions because we employed budget constraints rather than value or cost to determine the agent's maximum willingness to pay at any moment; therefore the intersection is not meant to predict the mean price.  

It was not necessary in this model for all buyers to meet their demand but it was convenient both for realism and reproducibility; alternative initial conditions that led to some buyers failing to meet their demand did not qualitatively change any of the results presented below.  It was enough for us to assume a $5\%$ excess total supply in order to ensure that all buyers in all trading sessions met their demand, i.e., none left the market because their budgets expired first.  Therefore only $5\%$ of the sellers failed to sell all of their supply (consistent with our choices for supply and demand) and only $3\%$ failed to meet their revenue goals (expressed in their budget) in all the trading sessions.  

The budget constraints in conjunction with an agent's initial endowment of multiple units of quantity played a central role in these bilateral transactions.  Each transaction began by randomly selecting a buyer and a seller from those remaining in the market.  The buyer entered her bid and the seller entered his ask for exactly one unit of power, independently of each other, as follows.  The buyer bid a random number drawn uniformly from the interval $(0, \frac{B}{D}]$, where $B$ was her remaining budget and $D$ was her remaining demand.  The seller's ask was determined from one of two cases: For a positive budget, the seller's ask would be a random number drawn uniformly from the interval $ [\frac{B}{S}, B]$, where $B$ was his remaining budget and $S$ was his remaining supply; otherwise, because we allowed negative seller budgets, he would ask a random number drawn uniformly from the interval $(0, \frac{B_0}{S_0}]$.  If the seller's budget were positive but he had only one unit of supply left, his ask would become exactly $B$; we chose budgets so that this case almost never occurred in practice.  Those few sellers with a positive budget at the end of the trading session were almost always those with many more than one unit remaining unsold.  Both of these choices for constructing bids or asks were conservative compared to the alternative (for the buyer) of bidding up to her entire budget for one unit or (for the seller) of always asking between zero and his whole budget.  That alternative would result in some buyers failing to meet their demand and more sellers failing to meet their revenue goals but it would not qualitatively change the results presented below.  (Another alternative would strongly impact the results, i.e., expressing the agent's preferences with fixed value or cost regardless of budget\cite{GodeSunder:1993}.  We note that in the case each agent is endowed with exactly one unit of quantity the budget constraint is also equivalent to fixed value or cost; nevertheless, whether the agents are endowed with one or many units, that alternative would not produce the funnel-shaped price divergence anticipated in Figure \ref{fig:Stamber}.)

The transaction succeeded if and only if the buyer's bid exceeded the seller's ask.  In that case, the sales price was determined by randomly distributing the surplus between the buyer and seller according to \begin{equation*}
price = (bid - ask)*\kappa + ask
\end{equation*} 
with $\kappa$ selected uniformly randomly on the unit interval (alternatively, one could have fixed $\kappa$ anywhere on the unit interval without qualitatively changing the results presented below).  Furthermore, each agent's budget was reduced by the sales price and each agent's quantity would be reduced by one unit.  The offers were presented as ``take it or leave it'', so that if the transaction failed, there would be no subsequent negotiation.  No agent employed the history of bids or asks to calculate future bids or asks.  There was no restriction on or charges for the number of transactions that were attempted; instead, buyers and sellers continued to be randomly paired until either there were no buyers left or no sellers left (it turned out that there were always sellers left because the buyers always met their demands with the parameters specified in \ref{appendix:initialization}).  A particular buyer-seller pair could be drawn randomly more than once because agents were sampled (with replacement) until an agent was removed from the market.  Transactions that were agreed upon were assumed to be feasible and free from transmission charges.  

Each attempted transaction, regardless of outcome, counted as one unitless step in the trading session.  The step plays the role of time only in the sense of imposing an ordering on the transactions.  In a real short-term power market, both buyers and sellers are in a race against the clock.  Here, the trading session was allowed to run as long as there was both supply and demand.  Therefore the time pressure on the agents manifested itself exclusively through the shrinking supply and demand; agents were not given clocks and could infer ``time'' only from their remaining budget and quantity.  

This model differs from the classical Edgeworth-box trading paradigm\cite{MICRO:2001a} in three key ways: (i) we abandoned the classical concept of bi-modal traders by instead fixing an agent as either a buyer or a seller (ii) we imposed a more restrictive specification of trading preferences (iii) we allowed at most one unit of power to be sold per transaction without any restriction on the other good (money).  In the classic two-good trading models without production, all traders would be assumed consumers of each good endowed with some initial quantity.  According to their indifference curves, they would attempt to make a trade which would increase (or at least not decrease) their utility.  There would be no restriction on the number of units sold for each good, except that they would be feasible according to the agent's endowments.  In making a trade, there would be no pre-determined buyer or seller; indeed a trader could switch between buying and selling as necessary to increase his utility.

Nevertheless, we imposed a more rigid structure for our model.  Buyers were identified and given a target quantity to buy; once it was obtained they were satisfied (never to buy more) and would not become sellers.  Likewise, sellers were given a target quantity to sell; once the supply was sold, they would not become a buyer.  Thus, we adopted the terms ``buyers'' or ``sellers'' to represent our trader agents since we restricted their behavior accordingly.  This rigid specification of buyer or seller preferences is not usually adopted in classical economics.  Typically preferences exhibit diminishing marginal utility and local nonsatiation, which were not adopted in this model.  Our buyer was not required to value the last unit of power any less than the first.  As a result, she may have paid a higher price for her last unit of good than she did for her first.  Additionally, ``more'' is not always ``better'' for the buyer, who in our model exits the market once her target demand is met.  Instead of allowing for substitutability between the two goods, we assumed that buyers were only interested in satisfying their target demand and sellers were only interested in selling their target supply given their respective budget and cost constraints. In both cases, these preferences are consequences of the impossibility of storing bulk power.  

\section{Results}
\label{sec:results}

\begin{figure}[p]
\begin{center}
\includegraphics[width=16cm]{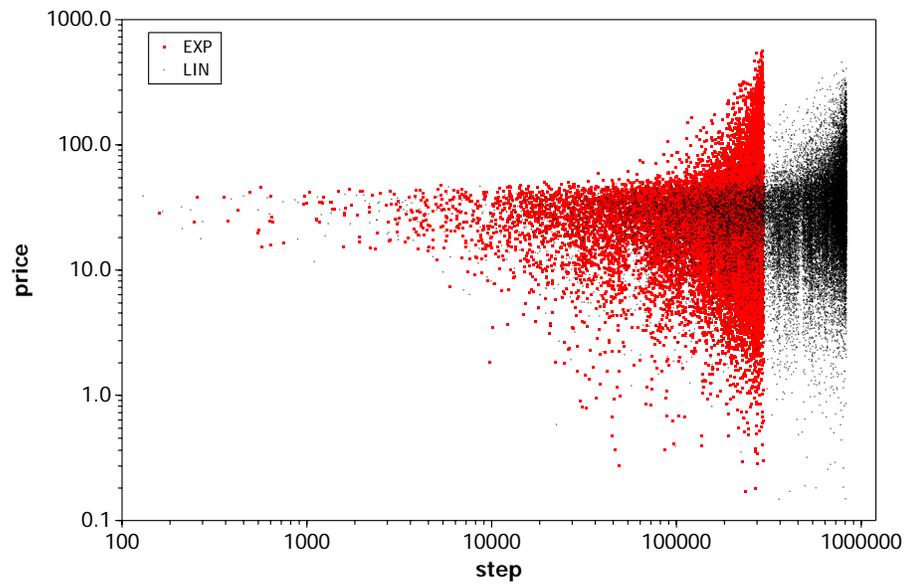}
\caption{Log-log plot of the sales price evolution generated in a typical trading session for cases EXP and LIN, respectively.}
\label{fig:log-log-evolution}
\end{center}
\end{figure}

For each of ten independent trading sessions we initialized two sets of buyers and sellers, one for each of the two cases EXP and LIN (see Appendix \ref{appendix:initialization} for the parameters).  In each set, each agent was assigned randomly selected budgets and quantities.  The two cases produced similar results.  Fig. \ref{fig:log-log-evolution} shows two typical trajectories of the sales prices for the two cases, respectively.  We note immediately the qualitative resemblance of both trajectories to the funnel shaped curve sketched in Fig. \ref{fig:Stamber} as the prices diverge both upward and downward away from the mean price in the latter part of the trajectories.  The principal differences between the two trajectories are that many more transactions were attempted in case LIN to close the market than in case EXP and that maximum prices in case LIN were lower than in case EXP.  For both trajectories, there was a period for which the prices fluctuate narrowly, corresponding to an elastic regime for both sellers and buyers, but the prices began to diverge after about $10^4$ steps, reaching their most extreme divergence near the end of the session.  

\begin{figure}[p]
\begin{center}
\includegraphics[width=16cm]{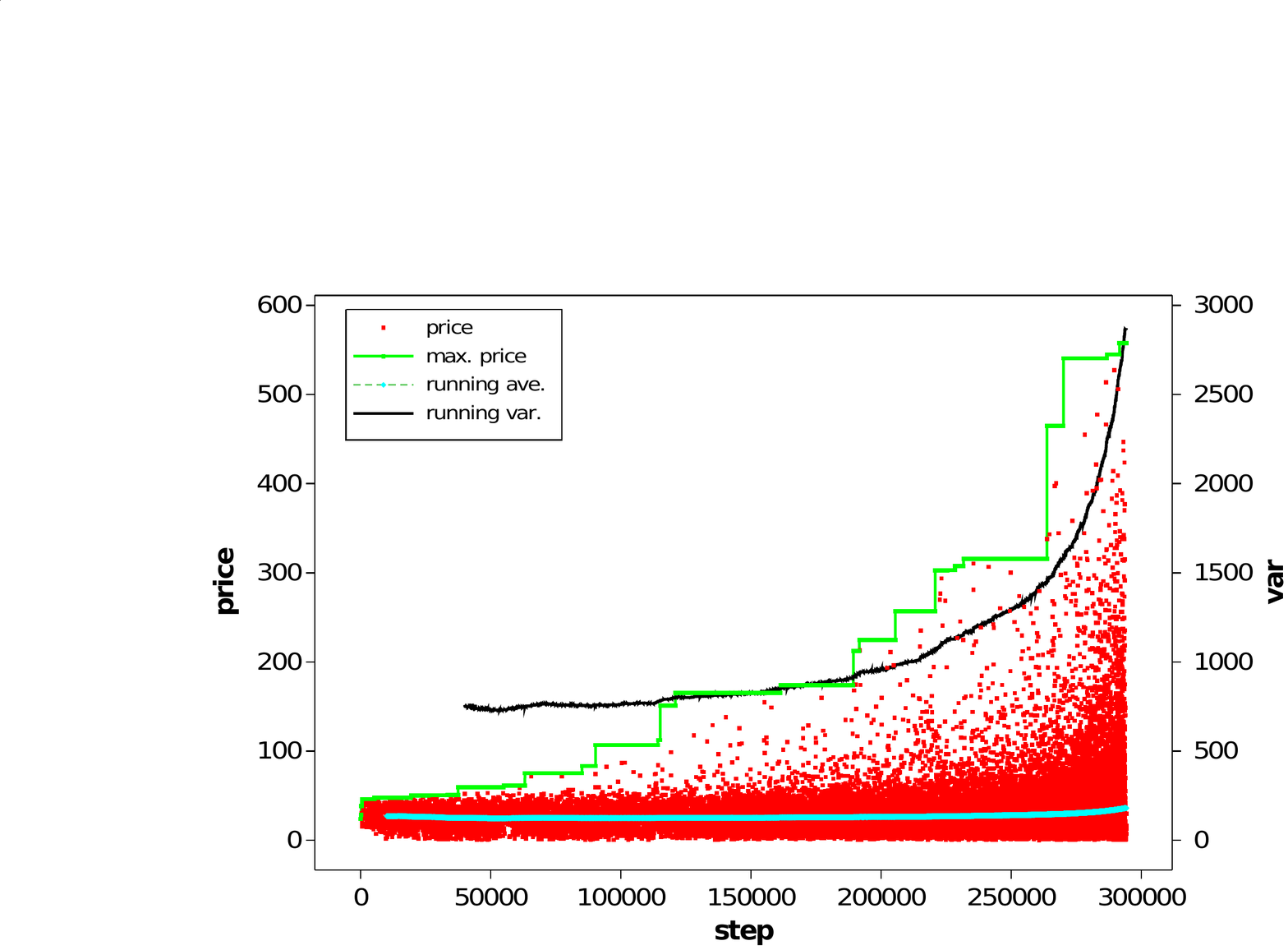}
\caption{Sales price evolution of a typical trading session (case EXP), shown also in Fig. \ref{fig:log-log-evolution}, in which the evolution of the maximum price is indicated.  The instantaneous price, the maximum price, the mean price (running mean) are read on the left-hand axis.  The variance (running variance) is read on the right-hand axis.}
\label{fig:linear-evolution}
\end{center}
\end{figure}

Fig. \ref{fig:linear-evolution} reproduces the typical trajectory for case EXP (but with a linear price scale) in order to show the typical rise of the maximum price, which was most dramatic as the market approached its close.  We also include in Figure 4 the running mean sales price and its variance, in order to show how they track the growth of the maximum price.  

The large divergences shown in Figs. \ref{fig:log-log-evolution} and \ref{fig:linear-evolution}, although anticipated (in fact required for a realistic treatment of this market), complicate the statistical analysis because the trajectories never settled upon a steady-state price.  Furthermore, the resulting price distributions were far from the normal, complicating the interpretation of statistics based on higher moments.  We note that the mean sales price (36 for EXP, 35 for LIN) did not vary much throughout the trajectories despite the large and rapidly growing variance, which in turn qualitatively tracked the growth of the maximum sales price.

\begin{figure}[p]
\begin{center}
\includegraphics[width=16cm]{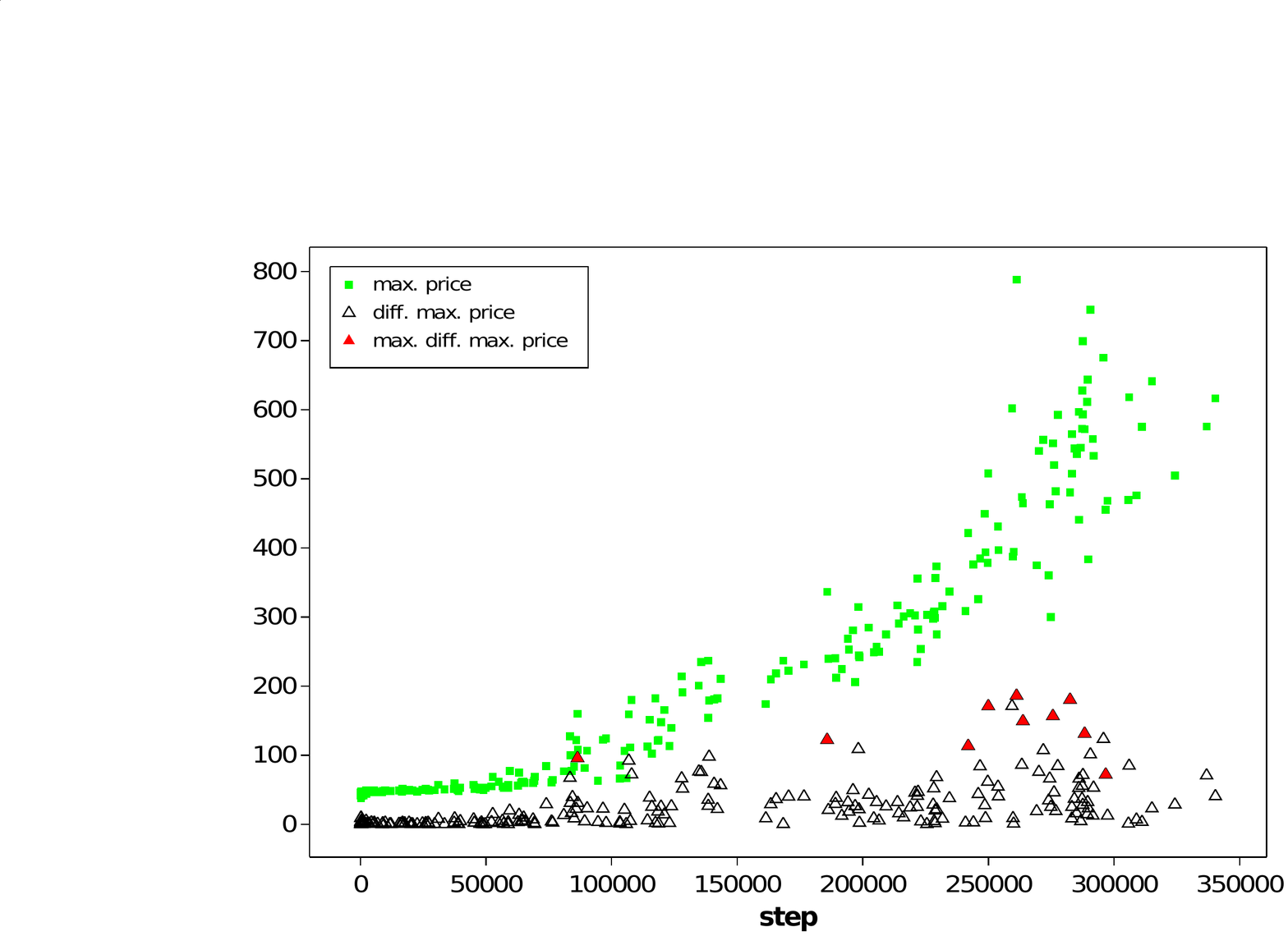}
\caption{Maximum price evolution of the ten trajectories (i.e., trading sessions).  The evolution of the change in the maximum price and the maximum (for each trajectory) of the change of maximum price are indicated with triangles.}
\label{fig:max}
\end{center}
\end{figure}

Fig. \ref{fig:max} shows the superposition of the ten trajectories of the maximum price of case EXP, along with the change in the maximum price.  We also note that large changes in the maximum price did not begin until after about the first one-fifth of each trajectory had evolved.  The maximum change in the maximum price for each trajectory (colored in red) occurred with one exception in the latter third of the trajectories.

\begin{figure}[p]
\begin{center}
\includegraphics[width=16cm]{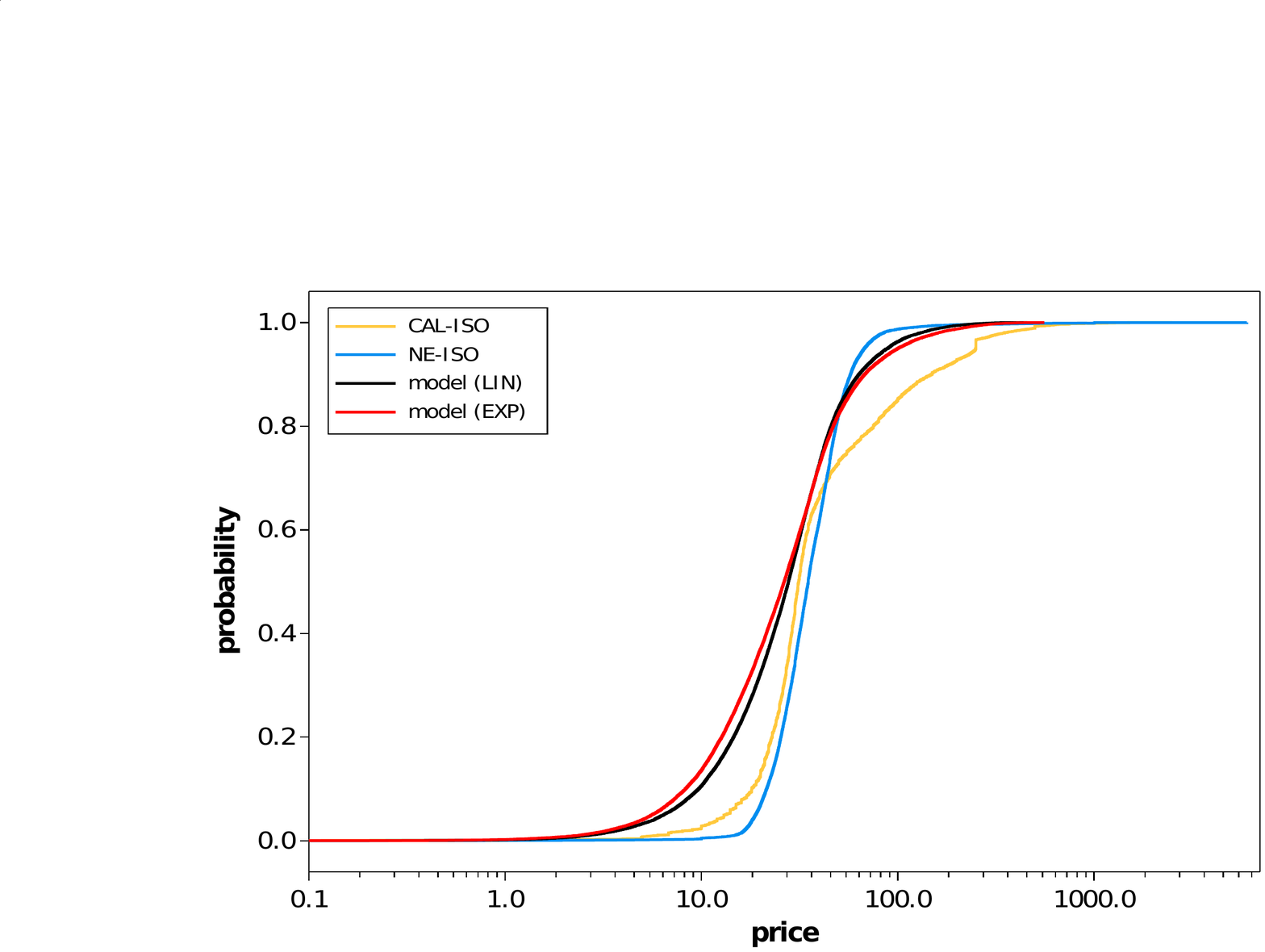}
\caption{Semi-log plot of the empirical cumulative distributions of the hourly sales price (USD/MWh) from NE-ISO and CAL-ISO compared with the distributions of the instantaneous (and unitless) sales price from the two cases of the model.}
\label{fig:semi-log-price}
\end{center}
\end{figure}

\begin{figure}[p]
\begin{center}
\includegraphics[width=16cm]{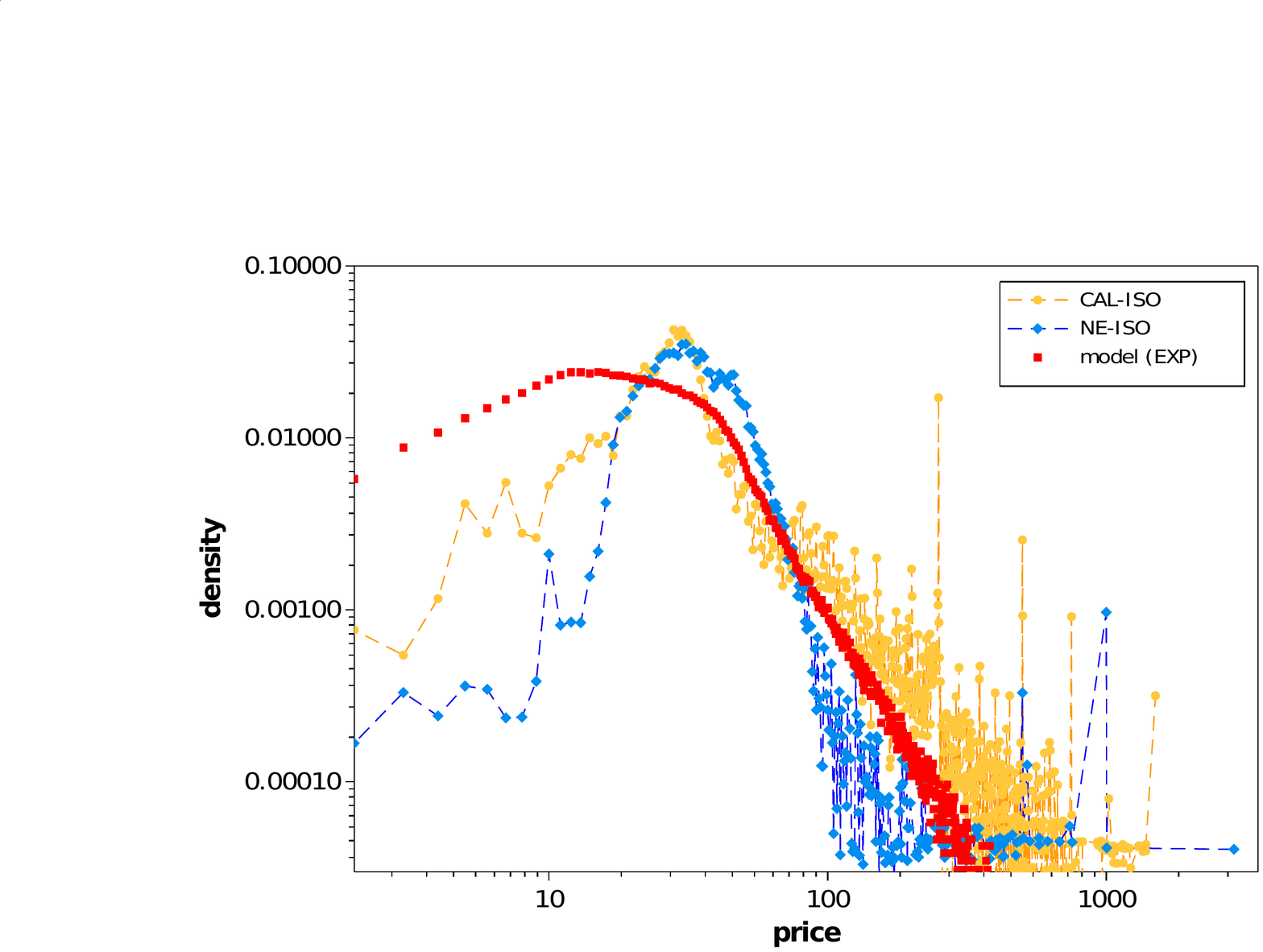}
\caption{Log-log plot of the empirical density distributions of the sales price.  The case LIN is omitted because it would be obscured (except for the lowest prices) by case EXP.}
\label{fig:log-log-price}
\end{center}
\end{figure}

The empirical probability functions (i.e., cumulative distribution f functions) of the sales prices from all ten trajectories are displayed in Fig. \ref{fig:semi-log-price}; the two cases nearly overlap.  We also displayed the empirical probability functions curves for the New England (NE-ISO) and California (CAL-ISO) sales price data (in USD/MWh); the NE-ISO data were collected from hourly reports for all of 1999-2002 and the CAL-ISO data were collected from April 1998 through January 2001.  The comparison of the model with the data is problematic, as we discuss below, because the price data results from hourly auctions instead of bilateral trades.  The S-shaped curve on a semi-log plot shows that all of the curves in Fig. \ref{fig:semi-log-price} resemble a log-normal distribution even though none of the curves strictly fit the log-normal.  The probability distributions displayed in Fig. \ref{fig:semi-log-price}  are fundamental, e.g., they do not require a prior choice for a bin size.  Nevertheless we display the density distribution function in Fig. \ref{fig:log-log-price}  (with a unit bin size) because the density may be more intuitive; in particular it is easier to see that the high price tail behavior is similar between the models and the data.  

\section{Discussion and Conclusions}
From inspection of the sales price trajectories themselves (Figs. \ref{fig:log-log-evolution} and \ref{fig:linear-evolution}) and the maximum prices in each of those trajectories (Figs. \ref{fig:linear-evolution} and \ref{fig:max}) we note that the sales prices diverge in the latter part of the trajectory from the mean price found in the early part of the trajectories.  The elasticity or inelasticity of the initial conditions (Figure 2) had little to do with the outcomes of the model or their comparison with data (Fig. \ref{fig:semi-log-price}).  With budget constraints for multiple units of supply or demand, the agents slowly but persistently adjusted their ability to pay according to their budgets so that the results were less sensitive to initial conditions with, e.g., preferences based on fixed values and costs, as in \cite{GodeSunder:1993}.  We note that the mean sales prices corresponding to the cases of elastic (LIN) and inelastic (EXP) initial conditions are similar to each other and that both are a little larger than the equilibrium price (about 30) that would be expected from a simple supply-demand curve analysis of the initial conditions; the increase was due (Fig. \ref{fig:linear-evolution}) almost entirely to the divergence of prices especially in the last third of the market trajectory.  

The comparison to the hourly sales price data is problematic because it records prices (USD/MWh) from auctions, not short-term bilateral trades; we would have compared with bilateral trade data if it had been publicly available.  Nevertheless the comparison is useful if for no other reason than to investigate the sensitivity of price divergence to market structure.  First, we note that the price distribution from data is narrower than the model.  This is expected when using bilateral trading versus auctions\cite{MICRO:2001a}.  Auctions and bilateral trades are two very different market structures; nevertheless the behavior in these two structures is similar for high prices.  In particular we note in Fig. \ref{fig:log-log-price} that the tail of the price distribution decreases at roughly the same rate for the model as for the data.  We made no attempt to fit or censor either the data or the results from the model displayed in Figures \ref{fig:semi-log-price} and  \ref{fig:log-log-price}; in particular, the CAL-ISO data includes prices from intervals when supply and demand were manipulated\cite{Sweeney:2002}.  

In our model the price divergence results from, on the one hand, the ability of some buyers to spend the remaining budget that resulted from making beneficial trades early in the market in order to obtain increasingly scarce supply to satisfy their demand and on the other hand, the ability of some sellers to sell their excess supply cheaply after they had achieved their revenue goals.  For both, the preference to either meet their entire demand, or sell their entire supply motivates each to continue trading even as prices diverge.  It may be that price divergence is not often observed in other markets because those agents may be more willing to refrain from trading altogether, especially if the quantity being traded is storable.  Our results suggest that price divergence is a phenomenon that is insensitive to the details of market structure and to the knowledge that agents might have about each other and of the market itself. In conclusion, the price divergences observed in short-term power markets that might have been due to wide range of causes, e.g., a deep knowledge of other traders, artifacts of market policies, or clever strategies by the agents, can also be generated in a model market of myopic agents with only simple budget constraints and preferences. 

\section*{Acknowledgements}
We thank Robert Axtell (George Mason), Verne Loose (RESPEC), and Paul Ormerod (Volterra) for helpful conversations. This work was supported in part by the Sandia National Laboratories Laboratory Directed Research and Development program, the Critical Infrastructure Protection/Decision Support Systems Program (CIP/DSS) and the National Infrastructure and Simulation Analysis Center (NISAC).  The CIP/DSS program was funded by the United States Department of Homeland Security's Science and Technology Directorate.  NISAC is a joint program at Sandia National Laboratories and Los Alamos National Laboratory, funded by the United States Department of Homeland Security's Preparedness Directorate, under the direction of the Infrastructure Protection/Risk Management Division.  Sandia National Laboratories is a multi-program laboratory operated by Sandia Corporation, a Lockheed Martin Company for the United States Department of Energy's National Nuclear Security Administration under contract DE-AC04-94AL85000.

\appendix
\section{Appendix: Bilateral trades in real power markets}
\label{appendix:bilateral}

The marketing of electric power resources for the purposes of maintaining system integrity while allowing for profit opportunities beyond the traditional limits of utility regulation have been in operation in the United States since the passage of the Public Utility Regulatory Policies Act of 1978 and further enhanced by language in the Energy Policy Act of 1992 encouraging wholesale power competition. Orders\cite{ORDER888,ORDER889} later issued by the Federal Energy Regulatory Commission (FERC) codified this language into useable rules.  Bilateral contracting among utilities served as one of a number of useful means for implementing these reforms.  

This trend grew in the last decade and a half with the implementation of market structures (following on the heels of the FERC Orders) which allowed for long-term contracting (typically on a bilateral basis) as well as short-term ``power pool'' markets designed for demand-gap-filling, typically centered on a Locational Marginal Pricing model, where bids to buy and sell power are settled by a central arbitrating body, and are based on the locations of supply and demand, transmission capacity constraints, and offered prices.  Markets for Ancillary Services (e.g., spinning reserves, reactive power) have also followed this central arbitration model.  The marketing of electric power and services has evolved from a wide variety of structures towards a Standardized Market Design (SMD), incorporating these market elements and others\cite{Kiesling:2002,Zhou:2003}.  We note that the SMD, as well as most currently developed energy markets, employ a combination of bilateral and central arbitration structures.  This reflects the desire of market designers to allow for markets to thrive while taking into consideration that the reliable supply of electric power is paramount.  Short-term settlements in a purely bilateral atmosphere might create opportunities for non-optimal solutions to the dispatch of power among market participants which, due to physical constraints of the operating system, could lead to an artificial lessening of reliability for the participants.  The central arbitration structure is designed so that demand is met in the short term at an equitable level of revenue or cost to the sellers or buyers, respectively, while ensuring that dispatch is resolved reliably.

Nonetheless, bilateral transactions serve the dominant share of marketed electricity\cite{Zhou:2003}.  Establishing bilateral transactions, both for the buyer and seller, minimizes risk.  For the seller, the inherent risk is that generation capacity which is unnecessary to meet local demand will go unused, and will not be needed in shorter-term markets due to market pressures from other sellers with similar exposures.  For the buyer, the risk is that unmet expected load, while able to be met with available capacity, will be done so only at an extraordinary premium.  For both the buyer and the seller, bilateral transactions minimize the risk that the agreed-to transaction might be subsequently infeasible because of transmission constraints (e.g., congestion).  The price element of this risk to buyer and seller alike is best expressed in the variance of the settlement price seen for electric power as a function of the time of settlement ex ante, as illustrated in Figure \ref{fig:Stamber} above.  Contracts placed well in advance of necessity are much more likely to be established at a value which has a minimal degree of variance from the long run marginal cost of operation of the facility (plus a small profit).  Contracts placed nearer to the time of necessity face high variability in general, and engender risk to buyer and seller alike.  Much of this is based on the level of demand relative to availability at the point of need.  In period of low demand relative to supply, the typical seller will be faced with taking any price, even if at an usually low price for the period, for the sake of operating the sold capacity.  In periods of high demand relative to supply, buyers are left with two alternatives: take whatever price is offered, or reduce demand (through planned outages, customer interruptions, and like actions).  This situation can be exacerbated by the introduction of bidding structures inherent in the market structures that create the opportunity for high settlement prices\cite{HockeyStick:2004}. Each of these transaction behaviors takes place with the full knowledge that most of the participants, both buyers and sellers, are profit-maximizing entities, responsible to shareholders, with all of the inherent risks\cite{Kiesling:2002}.

Bilateral contracting of short-term power (within the security requirements implemented in the designs of existing short-term markets) remains a useful structure, especially in transactions between and within areas of the North American power grid which have not yet implemented structures along the lines of the SMD.  Here, little time is left to waste, as agreements must be negotiated and transmission access rights secured in a limited window of opportunity.  Many of the markets following many of the aspects of the SMD have incorporated price caps into the structure; however, for those which have not (or have placed high cap values), and for those areas following bilateral practices, documented prices for energy have been seen at extraordinarily high levels, up two orders of magnitude of the price under typical operating condition, typically for small quantities of power over short periods of time, necessary to maintain system integrity through times of peak demand\cite{INDIANA:1998}.

\section{Appendix: Initialization of the agents}
\label{appendix:initialization}

Here we present details of the initialization of the agents.  We assigned each agent's initial quantity (supply $S_0$ or demand $D_0$) from the nearest integer of a random variate drawn uniformly from the interval $[lo, hi]$, where $lo$ and $hi$ are given in Table \ref{tbl:A} below.  Then we assigned each agent's budget first by drawing a random variate $X$ from the continuous probability distribution $F(x)$, where $x$ is a provisional cost or value, and 
\begin{equation}
F(x) = \frac{\exp(-c\cdot lo)-\exp(-c\cdot x)}{\exp(-c\cdot lo)-\exp(-c\cdot hi)}
\label{eq:B}
\end{equation}
with $x \in [lo,hi]$; both $x$ and $c$ are real. The initial budget $B_0$ for each agent was formed from the product of $X$ and the initial quantity; therefore the initial cost or value is interpreted as $\frac{B_0}{D_0}$ or $\frac{B_0}{S_0}$, respectively.  For small $|c|$, $F$ is essentially linear, giving rise to linear supply-demand curves (case LIN); otherwise, the supply-demand curve is essentially exponential (case EXP).  For the choice of parameters listed in Table  \ref{tbl:A} , supply and demand for case EXP were inelastic where the supply-demand curves crossed, while supply and demand for case LIN were much more elastic (see Figure 2) in the same region.  We created 10 sets of agents for each of the two cases, with $N_B = 1000$ buyers and $N_S = 500$ sellers for each set.  Reversing these numbers gave substantially the same results, as did choosing both to be the same. 

\begin{table}[h]
\caption{Parameters employed in initializing the agents for the two cases EXP and LIN (Equation \ref{eq:B}).}
\begin{center}
\begin{tabular}{|lr|r|r|}\hline
Quantity & &  EXP & LIN \\  \hline
 Supply & & &  \\ \hline
  & $lo$ & $10$ & $10$ \\ \hline
  & $hi$ & $100$ & $100$ \\ \hline
  Cost & & & \\ \hline
  & $lo$ & $10$ & $10$ \\ \hline
  & $hi$ & $200$ & $35$ \\ \hline
  & $c$ & $-0.15$ & $-0.01$ \\ \hline
  Demand & & & \\ \hline
  & $lo$ & $10$ & $10$ \\ \hline
  & $hi$ & $42$ & $42$ \\ \hline
  Value & & & \\ \hline
  & $lo$ & $5$ & $30$ \\  \hline
  & $hi$ & $50$ & $50$ \\ \hline
  & $c$ & $0.2$ & $0.01$ \\ \hline
\end{tabular}
\end{center}
\label{tbl:A}
\end{table}

\newpage
\bibliographystyle{alpha}
\bibliography{draft3}

\begin{thebibliography}{{Uni}96b}

\bibitem[GS93]{GodeSunder:1993}
D.~K. Gode and S.~Sunder.
\newblock Allocative efficiency of markets with zero-intelligence traders:
  Markets as a partial substitute for individual rationality.
\newblock {\em Journal of Political Economy}, 101:119--137, 1993.

\bibitem[HRO04]{HockeyStick:2004}
D.~Hurlbut, K.~Rogas, and S.~Oren.
\newblock Protecting the market from ``hockey stick'' pricing: How the {Public
  Utility Commission of Texas} is dealing with potential price gouging.
\newblock {\em The Electricity Journal}, 17:26--33, 2004.

\bibitem[{Ind}98]{INDIANA:1998}
{Indiana Utility Regulatory Commission}.
\newblock A review of the circumstances and factors which resulted in capacity
  shortages and price volatility in midwest electricity markets the week of
  {June 22, 1998}: Staff report, 1998.

\bibitem[KM02]{Kiesling:2002}
L.~Kiesling and B.~Mannix.
\newblock Standard market design in wholesale electricity markets: Can {FERC's}
  proposed structure adapt to the unknown?
\newblock Technical report, Reason Public Policy Institute, 2002.

\bibitem[PR01]{MICRO:2001a}
R.~S. Pindyck and D.~L. Rubinfeld.
\newblock {\em Microeconomics}.
\newblock Prentice-Hall, Upper Saddle River, NJ, fifth edition, 2001.

\bibitem[Sta01]{Stamber:2001}
K.~L. Stamber.
\newblock Understanding the power crisis: Infrastructure interdependencies
  issues.
\newblock Technical report, Sandia National Laboratories, NM, 2001.

\bibitem[Swe02]{Sweeney:2002}
J.~L. Sweeney.
\newblock {\em The California Electricity Crisis}.
\newblock Hoover Press, Stanford, CA, 2002.

\bibitem[{Uni}96a]{ORDER889}
{United States Federal Energy Regulatory Commission}.
\newblock Order 889: Open access same-time information system (formerly
  real-time information networks) and standards of conduct, 1996.

\bibitem[{Uni}96b]{ORDER888}
{United States Federal Energy Regulatory Commission}.
\newblock Order888: Promoting wholesale competition through open access
  non-discriminatory transmission services by public utilities; recovery of
  stranded costs by public utilities and transmitting utilities, 1996.

\bibitem[Zho03]{Zhou:2003}
S.~Zhou.
\newblock Comparison of market designs.
\newblock In D.~Hurlbut and R.~Greffe, editors, {\em Market Oversight Division
  Report}. Public Utility Commission of Texas, Austin, TX, 2003.

\end{thebibliography}
\end{document}